# Phononic Frequency Combs through Nonlinear Resonances

L. S. Cao[1,2,3,*], D. X. Qi[1], R. W. Peng[1,*], Mu Wang[1], and P. Schmelcher[2,3]

1) *National Laboratory of Solid State Microstructures and Department of Physics, Nanjing University, Nanjing 210093, China*

2) *Zentrum für Optische Quantentechnologien, Universität Hamburg, Luruper Chaussee 149, D-22761 Hamburg, Germany*

3) *The Hamburg Centre for Ultrafast Imaging, Luruper Chaussee 149,D-22761 Hamburg, Germany*

## Abstract

We explore an analogue of optical frequency combs in driven nonlinear phononic systems, and present a mechanism for generating phononic frequency combs through nonlinear resonances. In the underlying process, a set of phonon modes is simultaneously excited by the external driving which yields frequency combs with an array of discrete and equidistant spectral lines of each nonlinearly excited phonon mode. Frequency combs through nonlinear resonance of different orders are investigated, and in particular the possibility of correlation tailoring in higher order cases is revealed. We suggest that our results can be applied in various nonlinear acoustic processes, such as phonon harvesting, and can also be generalized to other nonlinear systems.

* Electronic addresses: lcao@physnet.uni-hamburg.de, rwpeng@nju.edu.cn

An optical frequency comb (FC) is a light source whose spectrum consists of a series of discrete, equally spaced elements. It has become an important coherent optical source with diverse applications, ranging from optical frequency metrology to ultracold gases [1,2]. The nonlinear parametric frequency conversion in microresonators of high resonance quality factor [3], manifests itself as a promising generating principle of FCs. In the underlying generating process, a continuous-wave (CW) laser is introduced to a microresonator, and is subsequently converted into the eigenmodes of the resonator with discrete, equidistant spectral lines. Despite the fact that phonons can be analogized to photons in many aspects, however, nonlinear dispersion relations of phonons prevent a direct analogue of optical FCs in phononic systems. In this letter we provide a novel mechanism for the generation of FCs independent of the dispersion relation, and demonstrate its existence in a phononic nonlinear system, *i.e.* the Fermi-Pasta-Ulam $\alpha$ (FPU-$\alpha$) chain.

The FPU-$\alpha$ ($\beta$) chains [4], containing triplet (quadruplet) nonlinear terms, manifest themselves as an ideal test bed of nonlinear phenomena, and the response of FPU-$\alpha$ ($\beta$) chains to external influence has been widely studied. The thermal equilibration and heat transport of chains coupled to a thermal reservoir [5,6,7], various nonlinear excitations, such as solitons [8,9], and the supertransmission [10,11] of FPU-$\alpha$ ($\beta$) chains under monochromatic driving are the representative phenomena of driven FPU systems. Nonlinear resonances (NRs) can also take place between external driving and intrinsic phonon modes in driven FPU systems, and the corresponding phonons are simultaneously excited by external driving, when the driving frequency matches the sum of the eigenfrequencies of a set of phonons. The nonlinear resonance serves as the source of FCs being proposed in this letter, and we will discuss the generation of FCs through NR of different orders. The FPU-$\alpha$ chain is a nonlinear three-wave-mixing system, and the first order NR refers to a pair of phonon modes simultaneously excited, which is termed as direct nonlinear resonance (DNR) hereafter. The second order NR, in which a triple pair of phonon modes are excited by a cluster of nonlinear terms, is termed as cluster nonlinear resonance (CNR). Once a phonon interacts strongly with atoms in a 1D nonlinear system, its energy is converted into two or more parts, and two or more additional phonons are excited. This process is well known for photons, but it remains uncommon for phonons [12,13]. Harvesting phonons may have potential applications on acoustic amplification by stimulated radiation (analogous to lasers), thermal information transfer, nonlinear acoustics, and even for the entanglement of phonons.

In this letter we consider an N-site FPU-$\alpha$ chain with a monochromatic driving force of strength $f$ and frequency $\omega_d$ applied to the first site, which manifests as a driving for all phonon modes. With fixed boundary conditions on both ends of the chain, the Hamiltonian can be written in the phonon mode picture [14], as:

$$H = \sum_{i=1}^{N} [\frac{P_i^2}{2} + \frac{\omega_i^2 Q_i^2}{2} + f_i \cos(\omega_d t) Q_i] + \sum_{i,j,k} \frac{\alpha B_{ijk}}{2(N+1)} Q_i Q_j Q_k \, . \quad (1)$$

The first term describes the linear phonon under monochromatic driving, and $Q_i(P_i)$ is the canonical coordinate (momentum) of the phonon, with $f_i = \sqrt{\frac{2}{N+1}} f \sin(\frac{\pi i}{N+1})$. The second term is the nonlinear phonon interaction term, with $B_{ijk} = (\delta_{i\pm j\pm k,0} - \delta_{i\pm j\pm k,2(N+1)})$, where all possible sign combinations are allowed. Dimensionless units are used, so the mass, the linear coupling constant and the spacing of atoms in the chain are set to unity. The phonon mode frequency is given as $\omega_i = 2\sin\left(\frac{\pi i}{2(N+1)}\right)$ for the fixed boundary conditions, while the effects discussed here are expected not to depend significantly on the choice of boundary conditions. We demonstrate here the existence and provide the physical properties of phononic FCs both analytically and numerically via the Poincaré–Lindstedt (PL) perturbation method [15] and time-integration of the classical equations of motion based on a Runge-Kutta method, respectively. In the numerical simulations FPU-α chains of 10 and 200 sites are considered.

*Frequency combs through DNR.* The DNR of a driven FPU-$\alpha$ chain refers to the excitation of a pair of phonon modes (PMs) $(Q_i, Q_j)$, when the frequency matching condition $\omega_d \approx \omega_i + \omega_j$ is fulfilled. Hereafter we denote PMs $(Q_i, ..., Q_l)$ as $Q_{i,...,l}$. $\omega_d$ is set beyond the dispersion band of the chain, in order to avoid the linear excitation of phonon modes close to $\omega_d$. In addition to the nonlinear dispersion relation, we also assume $Q_{i,j}$ to form the only pair of modes, which is in nonlinear resonance with the external driving.

Let us consider a general DNR of $Q_{i,j}$ and assume this pair is coupled by the nonlinear term $\frac{\alpha}{2(N+1)} Q_i Q_j Q_k$. Now the DNR can be understood within the truncated phase space spanned by the three $Q_{i,j,k}$. Applying the PL method in the truncated space [15], $Q_{i(j)}$ is obtained as

$$Q_{i(j)} = A_0^{i(j)} \cos\left(\tilde{\omega}_{i(j)} t\right) + \sum_{p \neq 0} A_p^{i(j)} \cos(\tilde{\omega}_{i(j)} + p\Delta\omega)t, \tag{2}$$

with the amplitude $A_p^{i(j)} \sim [\tilde{\omega}_{i(j)}^2 - \left(\tilde{\omega}_{i(j)} + p\Delta\omega\right)^2]^{-1}$. The solution (2) demonstrates that an array of equidistant spectral lines arises around the intrinsic frequency in the solutions of $Q_i$ and $Q_j$ with spacing $\Delta\omega = \omega_d - \tilde{\omega}_i - \tilde{\omega}_j$. These equidistant spectral lines manifest themselves as a new type of FC structure.

The generating process of FCs through DNR takes place when the summation mode of the external driving is suppressed by choosing the driving frequency beyond the dispersion band, and the driving can only excite eigenmodes $Q_i$ and $Q_j$ with $\omega_d \approx \omega_i + \omega_j$. On top of such a NR process, the frequency difference $\Delta\omega$ gives rise to a cascade-like generation of sub-spectral lines around $\tilde{\omega}_{i(j)}$ with equal distance $\Delta\omega$, *i.e.* the FCs. Comparing this to the existing generating mechanisms for FCs, such as parametric conversion in microcavities and higher harmonic generation in surface acoustic wave systems [16], FCs through DNR can greatly reduce the frequency separation between the spectral lines. This opens the possibility to improve precision in frequency metrology.

The FCs predicted by the PL method are confirmed by numerical simulations of a FPU-$\alpha$ chain of 10 sites, as shown in Figure 1. All numerical simulations are based on the complete Hamiltonian (1), and no truncation is applied in the simulations. The DNR of these two modes is quantified by their effective phonon number $N_{eff} = (P_i^2 + \omega_i^2 Q_i^2)/(2\omega_i)$, which is essentially the energy of the modes divided by the intrinsic frequency. The temporal evolutions of $N_{eff}$ of $Q_{1,10}$ are shown in Fig. 1(a), and identical trains of temporal pulses are observed for the two modes. Such pulse trains arise exclusively for $Q_{1,10}$, and manifest themselves as the temporal fingerprint of FCs, of which the pulse period is given by $T = 2\pi/\Delta\omega$. The FCs can be more directly seen in the spectra of $Q_{1,10}$, as shown in Fig. 1(b) and (c), respectively, where an array of equidistant spectral lines (the FC) is observed in each figure. The numerical simulation, in this way, provides clear temporal and spectral signature of the FCs. As also depicted in Figure 1, the DNR FCs of $Q_{1,10}$ represent a dominating process: the amplitudes of other PMs' are at least one order of magnitude smaller, thereby justifying the truncation of the small amplitude non-resonant modes, as applied in PL treatment.

It is important to study the dependence of $\tilde{\omega}_{i(j)}$ on $\omega_d$. Though it is difficult to analytically derive $\tilde{\omega}_{i(j)}$ in PL method, as this involves solving highly nonlinear equations, a qualitative understanding can be obtained by the equations for $\tilde{\omega}_{i(j)}$ with the lowest order renormalization,

$$\tilde{\omega}_{i(j)}^2 - \omega_{i(j)}^2 = -\frac{F_k^2 \tilde{\omega}_{j(i)}^2}{2\omega_{j(i)}^2[\tilde{\omega}_{j(i)}^2 - \left(\omega_d - \tilde{\omega}_{i(j)}\right)^2]}, \tag{3}$$

with $F_k = f_k/(\omega_k^2 - \omega_d^2)$. Without explicitly solving the equations, we can see that $\tilde{\omega}_{i(j)}$ strongly depends on $\omega_d$, and particularly the renormalization of the intrinsic frequency becomes larger when $\omega_d$ approaches the resonance, which effectively shifts the resonant point from $\omega_d$. There are various mechanisms of frequency normalization in nonlinear phononic systems, such as that induced by the external-field [17,18] and that induced by thermal equilibrium [19]. Different from these mechanisms, here the renormalization in equation (3) is induced by the NR foldover effect [20].

Figure 2 shows numerical simulations on the generation of FCs of $Q_{1,10}$. Since the dynamical behavior of the two modes is almost identical, here we show the results for $Q_1$ only. Figure 2(a) and (b) illustrate the dependences of FCs of $Q_1$ on driving frequency $\omega_d$ and driving force $f$, respectively. In Fig. 2(a), the evolution of the spectrum can be divided into three different regimes with respect to $\omega_d$, with two off-resonant and one resonant regime. In the off-resonant regimes of Fig. 2(a), two frequency branches, $\tilde{\omega}_1$ and $\omega_d - \tilde{\omega}_{10}$, can be identified, which resemble the PL solution up to the first-order perturbation theory. As approaching the resonant regime, a sharp transition to the comb structure occurs, and all the spectral lines become strongly dependent on $\omega_d$, indicating the renormalization effect of equation (3). Fig. 2(b) shows that the separation of the combs can be tuned by the strength of the external driving, and a linear-like dependence is observed, which indicates a high tunability of the phononic FCs. The phononic FCs via DNR require a weak external driving, and Fig. 2 shows that FCs via DNR arise in a narrow resonant frequency window over a relatively broad regime of the driving strength $f$.

*Frequency combs in CNR.* In an FPU-$\alpha$ chain, higher order nonlinear resonances can occur with simultaneous excitation of more than two phonon modes. Such NRs occur via a cluster coupling of multiple nonlinear terms, and is termed as cluster nonlinear resonance (CNR). The CNR resembles the resonance clustering in the evolutionary dispersive wave systems [21,22] and also the effective many-body interactions intermediated by virtual excitations [23]. We demonstrate here the generation of FCs via CNR of a triplet pair of phonons.

The CNR of a triplet pair $Q_{i,j,k}$ and the consequent generation of FCs can also be interpreted by PL method. The nonlinear resonance channel for a triplet pair with external driving is built by a series of nonlinear terms, for instance, the triplets $Q_{1,2,8}$ and $Q_{1,3,5}$ are coupled to the external driving by the cluster channels $(Q_1Q_1Q_2, Q_1Q_2Q_3, Q_2Q_3Q_5, Q_3Q_5Q_8)$ and $(Q_1Q_2Q_3, Q_2Q_3Q_5)$, respectively. Applying the PL method in the truncated space spanned by triplet and intermediate phonons leads to a FC solution for the triplet, as $Q_\alpha^0 = \sum_n A_{\alpha,n}^0 \cos(\tilde{\omega}_\alpha - n\Delta\omega_{ijk})t$, $(\Delta\omega_{ijk} = \omega_0 - \tilde{\omega}_i - \tilde{\omega}_j - \tilde{\omega}_k)$, with the amplitudes $A_{\alpha,n}^0 \sim [\tilde{\omega}_\alpha^2 - (\tilde{\omega}_\alpha - n\Delta\omega)^2]^{-1}$. At resonance $\Delta\omega_{ijk} \approx 0$, the amplitudes diverge and the CNR of $Q_{i,j,k}$ takes place with a series of spectral lines in the spectra of $Q_{i,j,k}$, and FCs arise in the spectra of the triplet phonons.

More importantly, the detailed properties of CNR channels give rise to additional terms in even higher order expansions, and modify the comb structure. Taking the CNR of $Q_{1,2,8}$ for instance, higher order expansions of PL method predict a satellite comb $Q_\alpha^1 = \sum_n A_{\alpha,n}^1 \cos(\tilde{\omega}_\alpha \mp \Delta\omega' - n\Delta\omega)t$ ( $\Delta\omega' = \tilde{\omega}_2 - 2\tilde{\omega}_1$, $\alpha \in \{1,2\}$ ), which are attributed to the nonlinear term $Q_1Q_1Q_2$ and provide additional correlation of $Q_1$ and $Q_2$. As this additional correlation is exclusively between $Q_1$ and $Q_2$, such satellite combs appear in the spectra of $Q_1$ and $Q_2$ only, but not $Q_8$. This example demonstrates the possibility of correlation tailoring in the CNR frequency comb generation by simply choosing triplet PMs pairs with particular correlations in the CNR channels.

Figure 3 shows numerical results for FCs of $Q_{1,2,8}$ and $Q_{1,3,5}$ generated via related CNRs. In Fig. 3(a), oscillations of $N_{eff}$ of $Q_{1,2,8}$ present a two-fold structure on different time scales. First, on the long time scale, a train of broad pulses appears for each mode, whereas a fast oscillation of large amplitude can be observed within each broad pulse of $Q_1$ and $Q_2$ that measures a short time scale. The inset of Fig. 3(a1) illustrates fast oscillations of $Q_1$ and $Q_2$ in a broad pulse, with a $\pi$-phase shift between the fast oscillations of the two PMs. Correspondingly, in the spectrum of each mode, there arises a main comb around the intrinsic frequency, and an additional satellite comb also appears in the spectra of $Q_1$ and $Q_2$. The temporal broad pulses of the three PMs are related to the main combs of these modes, and the fast oscillations of $Q_1$ and $Q_2$ within the broad pulse are attributed to the appearance of the satellite combs. The $\pi$-phase shift of the fast oscillations of $Q_1$ and $Q_2$ is given by the opposite location of the satellite combs with respect to the main combs. The above temporal and spectral properties are in agreement with the prediction of PL method.

In Fig. 3(b) we observe a train of broad pulses with relatively small fluctuation for the $N_{eff}$ oscillations of $Q_{1,3,5}$, and there is only one main comb in the spectrum of each PM, as shown in Fig. 3(b1)-(b3). The small fluctuation is induced by multiple intermediated couplings in the resonance channels. The difference of CNR of $Q_{1,3,5}$ and that of $Q_{1,2,8}$ is due to the absence of additional correlations in CNR channel of $Q_{1,3,5}$. Such differences in the CNR FCs demonstrate that the generating principle of FCs via nonlinear resonances can quantitatively tune the spacing and location, and qualitatively tailor the correlation as well.

The FCs generated through NR are robust with respect to intrinsic decoherent channels induced by nonlinear couplings between the NR PMs and the rest. To demonstrate this feature, we numerically investigate the DNR FC in an FPU-$\alpha$ chain with 200 sites, which possesses more decoherent channels than 10-site model. The results are shown in Fig. 4, where temporal and spectral fingerprint of FCs can be easily identified.

To conclude, we introduce here a new mechanism in generating FCs via nonlinear resonances. The universality of NRs in various nonlinear systems, ranging from mechanical cantilevers and chains of vibrators to nonlinear optics, suggests that the FCs via NRs occur in many nonlinear systems. In particular, the controllability of FCs via NRs provides a new approach in optimizing energy conversion rate in phonon harvesting. Moreover, NRs also exist in quantum nonlinear systems, e.g. quantum FPU chains, where $Q_i(P_i)$ become non-commuting operators, suggesting that it is possible to generalize FCs to quantum nonlinear systems and possess application in entanglement of phonons.

**Acknowledgement:**

L.C. would like to thank Andrey Matsko for inspiring discussions on phononic frequency combs, as well as Chang Guoqing and his colleagues for sharing knowledge on optical frequency combs. This work was supported by the MOST of China (Grants No. 2012CB921502 and 2010CB630705) and the NSFC (Grants No. 11034005, 11321063, 91321312). P.S. acknowledges financial support by the Deutsche Forschungsgemeinschaft (DFG) through the excellence cluster The Hamburg Centre for Ultrafast Imaging - Structure, Dynamics, and Control of Matter on the Atomic Scale.

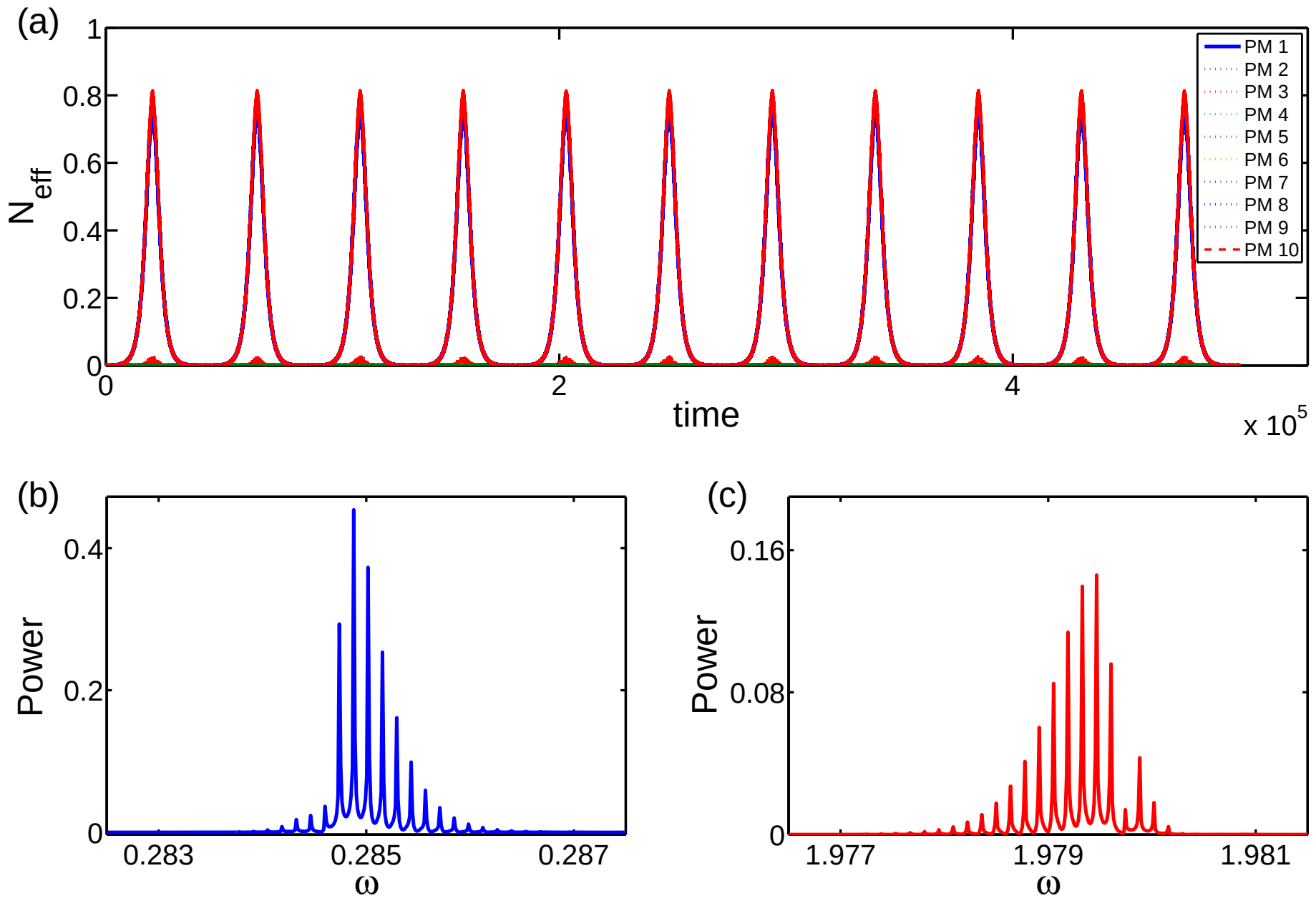

Figure 1. Illustrations of DNR FCs of $Q_{1,10}$ in a 10-site FPU-$\alpha$ chain, with $N_{eff}$ oscillations of all PMs shown in (a), and the spectra of $Q_{1,10}$ in (b) and (c), respectively. $N_{eff}$ oscillation of $Q_{1,10}$ overlap, and the oscillation amplitude of other PMs is at least one order of magnitude smaller than those of $Q_{1,10}$. Parameters are $(f,\alpha,\omega_d)=(0.1,0.1,2.264)$, and the frequency spectra are obtained by fast Fourier transformation of $Q_{1(10)}(t)$.

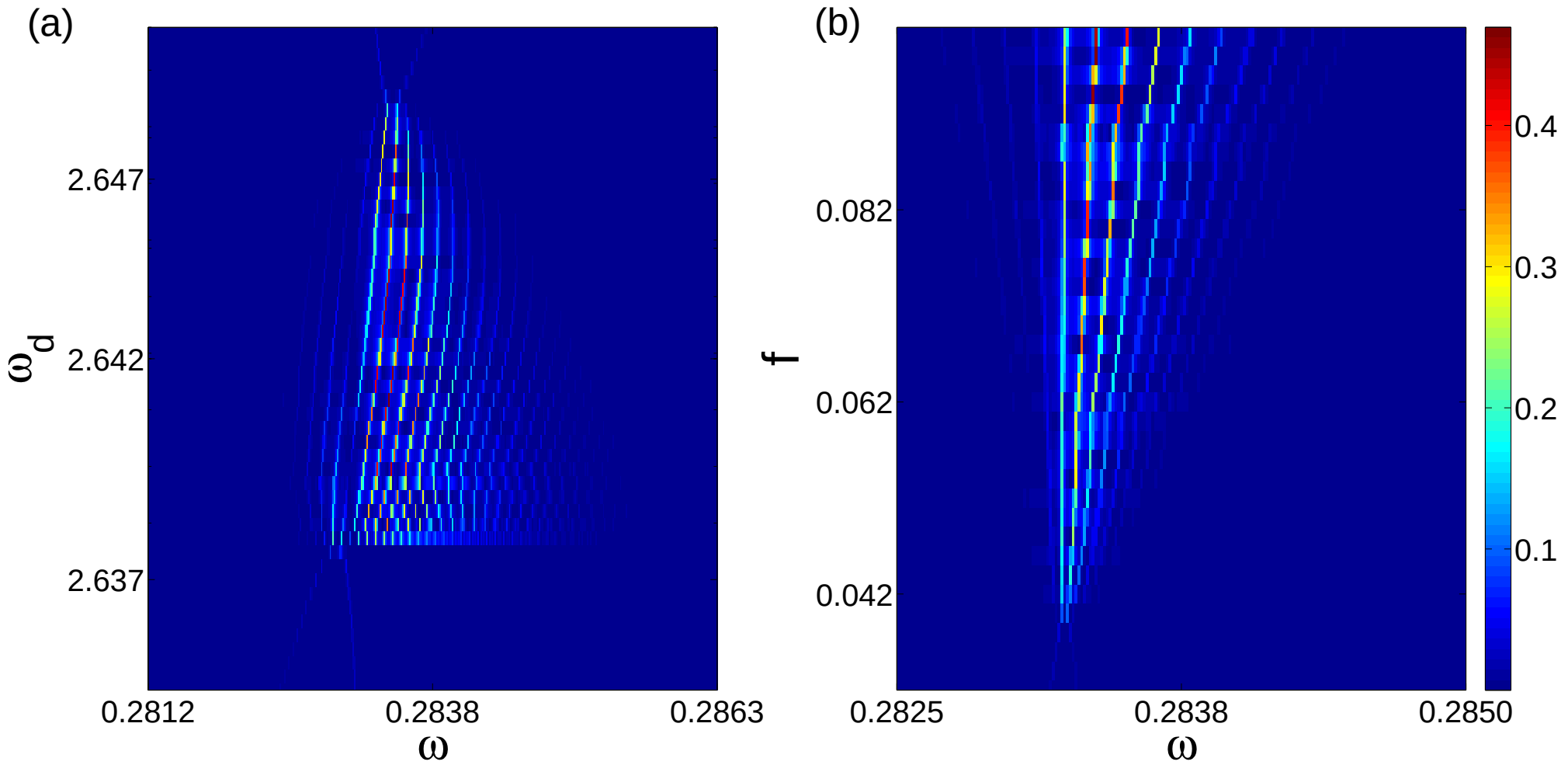

Figure 2. The frequency spectrum of $Q_1$ versus the external driving frequency $\omega_d$ (a) and the driving strength $f$ (b) in the vicinity of the DNR of $Q_{1,10}$, with the encoded color indicating the power of the spectrum.

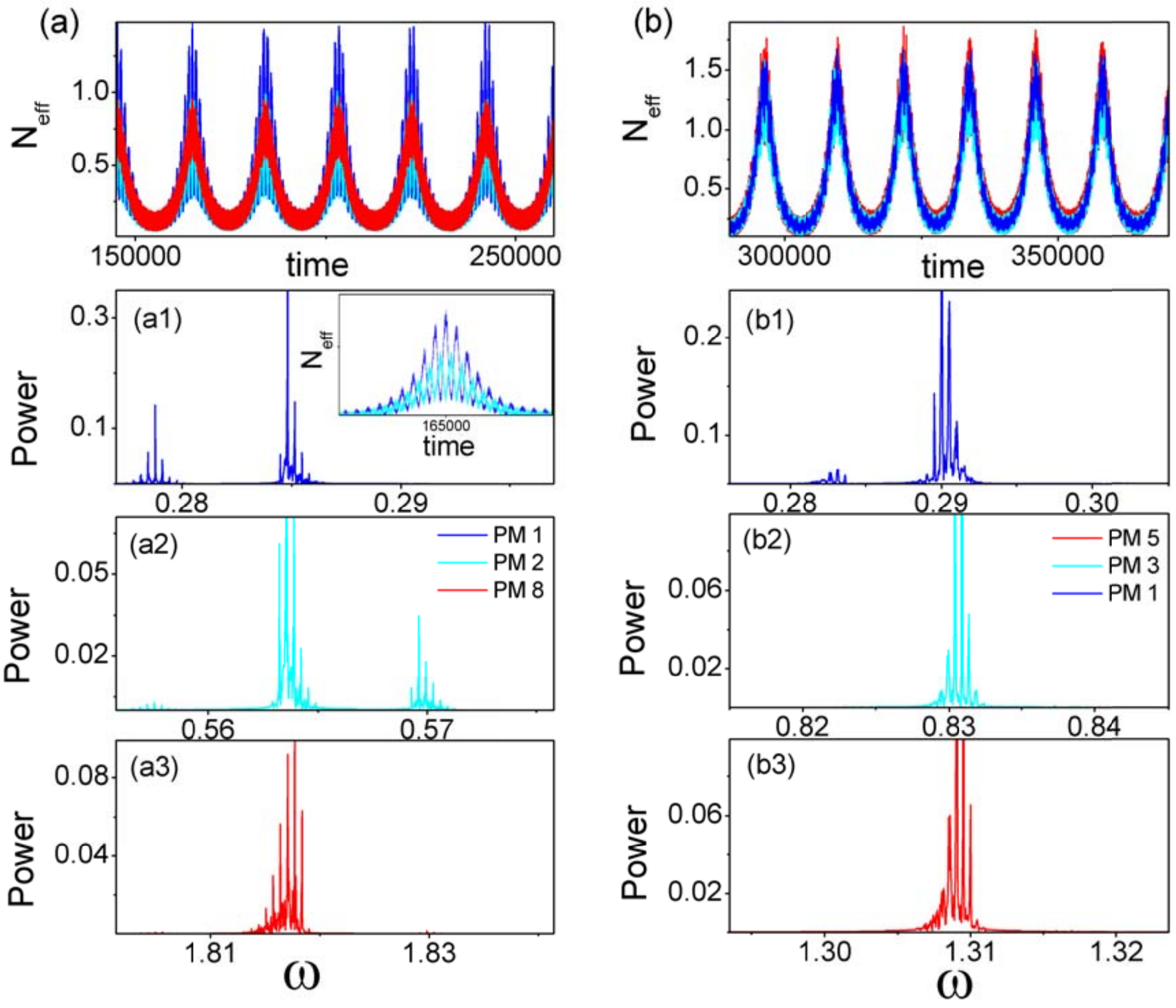

Figure 3. CNR FCs of $Q_{1,2,8}$ and $Q_{1,3,5}$. (a) and (a1)-(a3) show $N_{eff}$ oscillations of $Q_{1,2,8}$ and the spectra of corresponding PMs, respectively. (b) and (b1)-(b3) present $N_{eff}$ oscillations and corresponding spectra of $Q_{1,3,5}$, respectively. The inset of (a1) shows the detailed structure of $Q_1$ and $Q_2$ within one broad pulse.

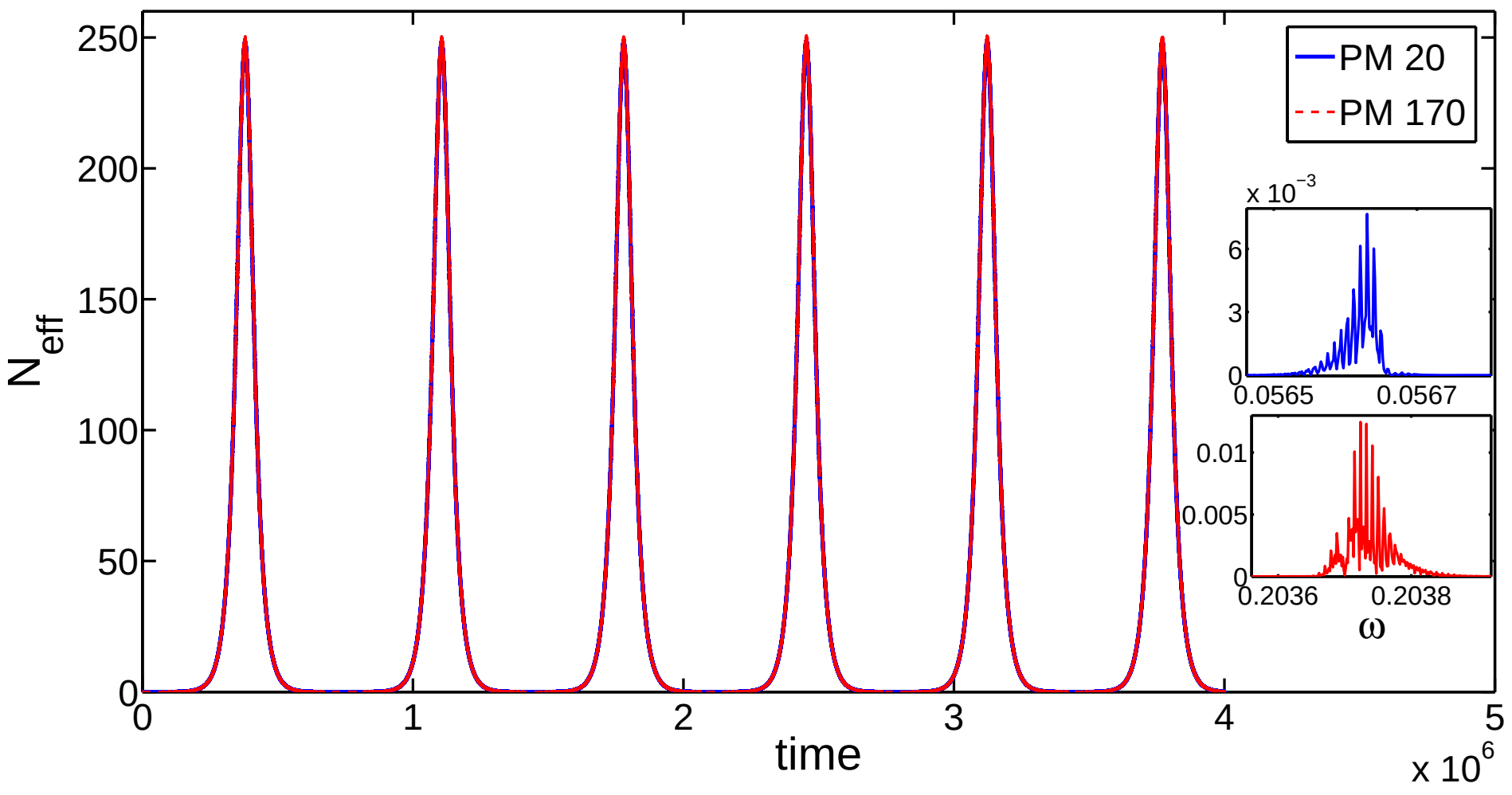

Figure 4. DNR FCs of $Q_{20,170}$ of an FPU-$\alpha$ chain of 200 sites. The main graph shows the oscillation $N_{eff}(t)$ of both PMs, and oscillation of the two modes lies on top of each other. The insets present the frequency spectra of $Q_{20,170}$, respectively.

# **Supplemental material:** Phononic Frequency Combs through Nonlinear Resonances

L. S. Cao[1,2,3,*], D. X. Qi[1], R. W. Peng[1,*], Mu Wang[1], and P. Schmelcher[2,3]

1) *National Laboratory of Solid State Microstructures and Department of Physics, Nanjing University, Nanjing 210093, China*

2) *Zentrum für Optische Quantentechnologien, Universität Hamburg, Luruper Chaussee 149, D-22761 Hamburg, Germany*

*3) The Hamburg Centre for Ultrafast Imaging, Luruper Chaussee 149,D-22761 Hamburg, Germany*

* Electronic addresses: lcao@physnet.uni-hamburg.de, rwpeng@nju.edu.cn

## I. Poincaré–Lindstedt method

The Poincaré–Lindstedt (PL) method is a widely used analytical tool for ordinary differential equations [1], with the ability to avoid long-time diverging terms, *i.e.* the secular terms, and we demonstrate here the use of the PL method in the driven FPU-$\alpha$ chain. The equations of motion for the PM $Q_i$ read:

$$\ddot{Q}_i = -\omega_i^2 Q_i + f_i \cos(\omega_d t) + \sum_{k,l} \frac{\alpha B_{ikl}}{2(N+1)} Q_k Q_l, \tag{1}$$

The PL method for a FPU-$\alpha$ chain mainly includes two steps: firstly in each equation (1), the time $t$ is rescaled to $\tau_i$ as $\tau_i = \tilde{\omega}_i t$, and the time derivative is consequently transferred to the derivative with respect to $\tau_i$ as $\partial_t^2 = \tilde{\omega}_i^2 \partial_{\tau_i}^2$. $Q_i$ and $\tilde{\omega}_i$ are then expanded according to $\alpha$ as $Q_i = \sum_{n=0} \alpha^n Q_i^{(n)}$ and $\tilde{\omega}_i^2 = \omega_i^2 + \sum_{n=1} \alpha^n \tilde{\omega}_{i,n}^2$. Substituting the expansions into equation (2) we obtain the differential equation for $Q_i^{(n)}$ with respect to $\tau_i$ as:

$$\begin{aligned} \partial_{\tau_i}^2 Q_i^{(n)} &= -Q_i^{(n)} + \frac{f_i}{\tilde{\omega}_i^2} \delta_{n,0} \cos(\frac{\omega_d}{\tilde{\omega}_i} \tau_i) \\ &+ \frac{1}{\tilde{\omega}_i^2} \sum_{l,k} \sum_{n_1+n_2=n-1} \frac{B_{ikl}}{2(N+1)} Q_l^{(n_1)} Q_k^{(n_2)} - \sum_{n_1+n_2=n} \frac{\tilde{\omega}_{i,n_1}^2}{\tilde{\omega}_i^2} \partial_{\tau_i}^2 Q_i^{(n_2)}. \end{aligned} \tag{2}$$

Equation (2) can be solved sequentially from $n = 0$ to higher order. In the PL method, the divergence problem of normal perturbation theory is avoided by the last term of equation (2), where $\tilde{\omega}_{i,n}^2$ is determined by the request of the cancellation of the diverging term appearing in $Q_l^{(n_1)} Q_k^{(n_2)}$, when previous order terms are exploited. In this way we adapt the PL method to the driven nonlinear FPU-α chain.

## II. Analytical solution of DNR frequency combs via PL method

We consider a truncated phase space spanned by three PMs $(Q_i, Q_j, Q_k)$, which are coupled by nonlinear term $\frac{\alpha}{2(N+1)}Q_iQ_jQ_k$. The equation of motion for $Q_a$ ($a=i,j,k$) in the truncated space is given by:

$$\ddot{Q}_a = -\omega_a^2 Q_a + f_a\cos(\omega_d t) + \frac{\alpha}{2(N+1)}Q_bQ_c, \tag{3}$$

with $(a,b,c)$ belonging to $\{(i,j,k),(j,k,i),(k,i,j)\}$.

To apply PL method, firstly we expand the phonon canonical coordinates and corresponding renormalized eigenfrequency according to $\alpha$, as $Q_a = \sum_{n=0}\alpha^n Q_a^{(n)}$ and $\tilde{\omega}_a^2 = \omega_a^2 + \sum_{n=1}\alpha^n\tilde{\omega}_{a,n}^2$, and further we rescale the time as $\tau_a = \tilde{\omega}_a t$. Then the differential equation for $Q_a$ ($a=i,j,k$) becomes:

$$\begin{aligned}\partial_{\tau_a}^2 Q_a^{(n)} &= -Q_a^{(n)} + \frac{f_a}{\tilde{\omega}_a^2}\delta_{n,0}\cos(\frac{\omega_d}{\tilde{\omega}_a}\tau_a)\\ &\quad+\frac{1}{\tilde{\omega}_a^2}\sum_{n_1+n_2=n-1}\frac{\alpha}{2(N+1)}Q_b^{(n_1)}Q_c^{(n_2)} - \sum_{n_1+n_2=n}\frac{\tilde{\omega}_{a,n_1}^2}{\tilde{\omega}_a^2}\partial_{\tau_i}^2 Q_a^{(n_2)},\end{aligned} \tag{4}$$

Solving equation (4) order by order, $Q_a^{(0)}$ is just the linear solution of equation (3) with $\alpha=0$, while the solution $Q_{i(j)}^{(1)}$ (denoting both $Q_i^{(1)}$ and $Q_j^{(1)}$) with leading terms $\cos(\tilde{\omega}_{i(j)}t)-\cos\left(\omega_d-\tilde{\omega}_{j(i)}\right)t$ of amplitude $\left[\tilde{\omega}_{i(j)}^2-\left(\omega_d-\tilde{\omega}_{j(i)}\right)^2\right]^{-1}$, gives the frequency matching condition of the nonlinear resonance. Since $Q_k$ mainly plays the role of intermediating the nonlinear coupling between the external driving and PMs $(Q_i,Q_j)$ and is not significantly excited, it is justified to take into account only $Q_k^{(0)}$ in the solution.

Solving sequentially equation (4) for $Q_{i(j)}^{(n)}$ for higher orders gives rise to a series of near-resonant terms of $\cos(\tilde{\omega}_{i(j)}-p\Delta\omega)t$ ($p\in Z$) in the solution of $Q_{i(j)}$, with $\Delta\omega = \omega_d - \tilde{\omega}_i - \tilde{\omega}_j$. Keeping only the leading terms of $Q_{i(j)}^{(n)}$, the general solution of $Q_{i(j)}$ is given as:

$$Q_{i(j)} = A_0^{i(j)} \cos\left(\tilde{\omega}_{i(j)} t\right) + \sum_{p \neq 0} A_p^{i(j)} \cos(\tilde{\omega}_{i(j)} + p\Delta\omega)t, \tag{5}$$

with the amplitude $A_p^{i(j)} \sim [\tilde{\omega}_{i(j)}^2 - \left(\tilde{\omega}_{i(j)} + p\Delta\omega\right)^2]^{-1}$. Terms belonging to larger indices $p$ stem from higher order perturbation expansions. The solution (5) demonstrates that an array of equidistant spectral lines arises around the intrinsic frequency in the solutions of $Q_i$ and $Q_j$ with the spacing $\Delta\omega = \omega_d - \tilde{\omega}_i - \tilde{\omega}_j$. These equidistant spectral lines manifest themselves as a new type of FC structure.

## Ref: